\documentclass[a4paper,fleqn]{cas-dc}

\usepackage[numbers,sort&compress]{natbib}
 
\def\tsc#1{\csdef{#1}{\textsc{\lowercase{#1}}\xspace}}
\tsc{WGM}
\tsc{QE}

\begin{document}
\let\WriteBookmarks\relax
\def\floatpagepagefraction{1}
\def\textpagefraction{.001}

\shorttitle{}    

\shortauthors{}  

\title [mode = title]{Growth of hexagonal BN crystals by traveling-solvent floating zone}  



%

\author[1]{Eli Zoghlin}[orcid = 0000-0002-8160-584X]
\cormark[1]
\cortext[1]{Corresponding author}
\ead{ezoghli1@jh.edu}


\credit{Conceptualization, Methodology, Formal analysis, Investigation, Writing - Original Draft, Visualization}
\affiliation[1]{organization={William H. Miller III Department of Physics and Astronomy, The Johns Hopkins University},
            city={Baltimore},
           citysep={}, 
            postcode={21218}, 
            state={Maryland},
            country={United States}}

\author[2]{Juliette Plo}
\credit{Investigation}
\affiliation[2]{organization={Laboratoire Charles Coulomb, Universit{\'e} de Montpellier and CNRS},
            city={Montpellier},
           citysep={}, 
            postcode={34095}, 
            country={France}}

\author[3]{Gaihua Ye}[orcid = 0009-0005-2791-9636]
\credit{Investigation}
\affiliation[3]{organization={Department of Electrical and Computer Engineering, Texas Tech University},
            city={Lubbock},
           citysep={}, 
            postcode={79409}, 
	state = {Texas},
            country={United States}}

\author[3]{Cynthia Nnokwe}
\credit{Investigation}

\author[4]{Reina Gomez}[orcid = 0000-0002-2190-600X]
\credit{Investigation}
\affiliation[4]{organization={Materials Department, University of California, Santa Barbara},
            city={Santa Barbara},
           citysep={}, 
            postcode={93106}, 
	state = {California},
            country={United States}}

\author[5]{Austin Ferrenti}[orcid = 0000-0003-2075-9041]
\credit{Investigation,Writing - Review \& Editing}
\affiliation[5]{organization={Department of Chemistry The Johns Hopkins University},
            city={Baltimore},
           citysep={}, 
            postcode=21218, 
	state = {Maryland},
            country={United States}}

\author[1,5]{Satya Kushwaha}[orcid = 0000-0002-3169-969X]
\credit{Conceptualization, Methodology,Writing - Review \& Editing}

\author[3]{Rui He}[orcid = 0000-0002-2368-7269]
\credit{Resources, Writing - Review \& Editing, Supervision, Funding acquisition}

\author[4]{Stephen D. Wilson}[orcid = 0000-0003-3733-930X]
\credit{Conceptualization, Resources, Writing - Review \& Editing, Supervision, Funding acquisition}

\author[2]{Pierre Valvin}[orcid = 0000-0002-2356-2589]
\credit{Investigation, Validation}

\author[2]{Bernard Gil}[orcid = 0000-0002-1588-887X]
\credit{Validation}

\author[2,6]{Guillaume Cassabois}[orcid = 0000-0001-5997-4609]
\credit{Resources, Writing - Review \& editing, Supervision, Funding acquisition}
\affiliation[6]{organization={Institut Universitaire de France},
            city={Paris},
           citysep={}, 
            postcode={75231}, 
            country={France}}

\author[7]{James H. Edgar}[orcid = 0000-0003-0918-5964]
\credit{Conceptualization, Writing - Review \& Editing, Funding acquisition}
\affiliation[7]{organization={Tim Taylor Department of Chemical Engineering, Kansas State University},
            city={Manhattan},
           citysep={}, 
            postcode={66506}, 
	state = {Kansas},
            country={United States}}

\author[1,5,8]{Tyrel M. McQueen}[orcid = 0000-0002-8493-4630]
\ead{mcqueen@jhu.edu}
\cormark[1]
\credit{Conceptualization, Methodology, Writing - Review \& Editing, Supervision Funding acquisition}
\affiliation[8]{organization={Department of Materials Science and Engineering, The Johns Hopkins University},
            city={Baltimore},
           citysep={}, 
            postcode={21218}, 
	state = {Maryland},
            country={United States}}

\begin{abstract}
Large, high-purity single-crystals of hexagonal BN (h-BN) are essential for exploiting its many desirable and interesting properties. Here, we demonstrate via X-ray tomography, X-ray diffraction and scanning electron microscopy that h-BN crystals can be grown by traveling-solvent floating-zone (TSFZ). The diameters of grown boules range from 3 -- 5 mm with lengths from 2 -- 10 mm. Tomography indicates variable grain sizes within the boules, with the largest having areas of $\approx$ 1 mm $\times$ 2 mm and thickness $\approx$ 0.5 mm. Although the boules contain macroscale flux inclusions, the h-BN lattice itself is of high quality for samples grown under optimized conditions. The currently optimized growth procedure employs an Fe flux, moderate N$_2$ pressure ($P_{N2} \approx$ 6 bar), and a growth rate of 0.1 mm/h. Raman spectroscopy for an optimized sample gives an average linewidth of 7.7(2) cm$^{-1}$ for the E$_{\mathrm{2g}}$  intralayer mode at 1365.46(4) cm$^{-1}$  and 1.0(1) cm$^{-1}$ for the E$_{\mathrm{2g}}$  interlayer shear mode at 51.78(9) cm$^{-1}$. The corresponding photoluminescence spectrum shows sharp phonon-assisted free exciton peaks and minimal signal in the energy range corresponding to carbon-related defects ($E$ = 3.9 -- 4.1 eV).  Our work demonstrates the viability of growing h-BN by the TSFZ technique, thereby opening a new route towards larger, high-quality crystals and advancing the state of h-BN related research. 
\end{abstract}





\maketitle



\section{Introduction}\label{intro}

The utility of hexagonal BN (h-BN) has been recognized since at least the 1950’s, particularly for high-temperature applications, where its chemical inertness, thermodynamic stability and high electrical resistivity are extremely useful. The arrival of graphene \cite{novoselov2004,novoselov2005} subsequently stimulated additional interest in h-BN as an ideal insulating substrate for device applications \cite{dean2010}. h-BN is an excellent substrate for graphene-based devices due to the combination of a small lattice mismatch and atomic-level flatness (i.e. no dangling bonds) with good dielectric \cite{novoselov2004,novoselov2005,cassabois2016} and thermal transport \cite{yuan2019,zhang2017} properties. The chemical inertness and oxidation resistance of h-BN \cite{liu2013} also make it valuable for capping and encapsulation purposes \cite{siskins2019}. Furthermore, h-BN can also take an active role, as in graphene/h-BN moiré superlattices \cite{chen2019a,chen2019b,shi2014,ponomarenko2013}. The same properties which make h-BN ideal for graphene also make it a useful platform for device creation using other two-dimensional (2D) materials and it has seen widespread deployment in this context. 

More recently, h-BN is under study for a host of additional applications, including, among others, photonics, flexible electronics, non-linear optics and quantum information science and sensing technologies \cite{gil2020,roy2021,izyumskaya2017,vaidya2023}. A number of these applications exploit the intrinsic properties of h-BN on their own. As a naturally hyperbolic \footnote{Here, hyperbolic indicates that the relative permittivity changes sign as well as magnitude along different crystallographic directions.} birefringent material in the mid-infrared range h-BN is well-suited to many photonics applications \cite{caldwell2019}, such as confinement and manipulation of phonon polaritons \cite{dai2014}. Relatedly, point defects within h-BN have also begun to generate significant interest as single photon emitters \cite{tran2016,bourrellier2016,fournier2021}, which have the potential to serve as an optical platform for quantum information \cite{lounis2005} and as a spin defect for use in atomic-resolution quantum sensing \cite{bradac2021,lyu2022,udvarhelyi2023}.

Many of the aforementioned advances and discoveries were enabled by progress in the synthesis of high-quality, crystalline h-BN. Production of such samples as thin-films remains an ongoing effort and a number of techniques have been explored, particularly chemical vapor deposition \cite{dahal2011,ma2022,wang2024,chen2020,fukamachi2023,moret2021} and molecular beam epitaxy \cite{page2019,tsai2009,vuong2017,cheng2018,laleyan2018}. While work in this area is essential since there is potential to create wafer-scale single crystals which can be integrated into existing industrial practices, these methods typically lead to samples with relatively poor crystallinity, a high density of structural defects, and significant lattice strain due to mismatch with the substrate. This disorder is reflected in, for example, the linewidths of Raman peaks and in the lack of phonon-assisted free exciton signals in the photoluminescence spectrum. Furthermore, though progress has been made \cite{alemoush2024}, the thickness of the films is limited. 

Accordingly, the primary source of high-quality h-BN single crystals for the most demanding experiments has principally been the crucible-based flux (CBF) method developed by Taniguchi and Watanabe \cite{taniguchi2007}, which utilizes a Ba$_{3}$BN$_{2}$ flux in Mo crucibles at high temperature and GPa-scale pressures (commonly referred to as the high-pressure, high-temperature, or HPHT, technique). However, subsequent work has demonstrated that single-crystal h-BN samples of comparable quality and size can be made with various transition metal based fluxes such Fe-Cr or Ni-Cr alloys \cite{kubota2008,liu2017,liu2018,zhang2019,li2020b,li2020a,li2021} (referred to as the atmospheric-pressure, high temperature, or APHT, technique). These crystals have been utilized successfully in, among others, a number of interesting photonics experiments \cite{li2018,giles2018}. Despite these accomplishments, it is clear that further improvements are needed to enable the most ambitious technological proposals, particularly in 2D materials-based heterostructures and quantum information science related arenas. For example, charged impurities in h-BN likely represent the limiting source of disorder in graphene/h-BN heterostructures and are detrimental to any attempts at manipulation of charged quasiparticles for quantum information science applications \cite{kitaev2006,kitaev2003,rhodes2019}. 

From a practical perspective, these transition metal fluxes represent a substantive synthetic advance due to their ability to dissolve and recrystallize high-quality h-BN at atmospheric pressure. This greatly reduces the technical difficulty of producing single crystals via the CBF technique, enabling production without the highly specialized equipment needed for the Ba$_{3}$BN$_{2}$ flux \cite{taniguchi2007}. An as yet unrealized benefit of these fluxes is the possibility of applying the traveling-solvent floating-zone (TSFZ) method to growth of h-BN crystals. This is particularly exciting because the nature of the TSFZ method makes it well-suited to overcoming the main synthetic challenge associated with fully realizing the technological possibilities of h-BN: producing high-volume crystals with a low defect density. As a form of continuous, directional crystallization, the TSFZ method is a route towards larger, bulk, single-crystals than what is typically produced by the CBF method – which are generally on the order of 1 mm in lateral dimensions and up to 100 $\mu$m thick – while retaining the high quality. For one, TSFZ functions intrinsically with very high temperature gradients and experiments with CBF-growth under a temperature gradient have demonstrated improved structural quality as compared to slow cooling alone \cite{li2020b}. 

The TSFZ method also has a number of benefits with respect to sample purity which can help to reduce the extrinsic contamination seen in CBF-grown crystals via e.g. cathodoluminescence and secondary ion mass spectrometry \cite{onodera2019}. First, TSFZ is a "crucible-free" growth technique in the sense that the liquid flux does not come into contact with any additional material besides the one being grown. Second, TSFZ benefits from the zone-refining phenomenon, wherein impurities are retained in the molten zone due to a higher liquid solubility than in the rcrystallized solid \cite{pfann1957}. 

In this work, we report the feasibility of the TSFZ technique for crystal growth of h-BN. The effects of a variety of TSFZ growth parameters were studied, guided by efficient first-pass characterization via X-ray tomography. Two variations of TSFZ proved to be feasible for growing h-BN crystals and the grown boules range in size, with diameters of 3 -- 5 mm and lengths from 2 -- 10 mm. Tomography data indicates the presence of grains of variable size within the boules, with the largest observed having an area of $\approx$ 1 × 2 mm and thickness $\approx$ 0.5 mm. For optimized growth conditions (TSFZ geometry, Fe flux, $P$ = 7 bar, 0.1 mm/h growth rate) average Raman linewidths of 7.7(2) cm$^{-1}$ and 1.0(1) cm$^{-1}$ were obtained for the E$_{\mathrm{2g}}$ modes at 1365.46(4) cm$^{-1}$ and 51.78(9) cm$^{-1}$, respectively. The corresponding PL spectrum shows sharp phonon-assisted free exciton peaks as well as minimal signal in the energy range corresponding to carbon defects ($E$ = 3.9 -- 4.1 eV). 

Our paper is organized as follows. First, an outline of the crystal growth methodology is presented – describing the two geometries utilized and the various growth parameters – followed by descriptions of the characterization methodologies. Next, the results of various growth attempts and their characterization are presented. We then discuss the effects of the different growth parameters and summarize the current optimized growth procedure. Finally, a comparison of the highest quality TSFZ sample to CBF samples is presented. We conclude with a brief summary of future avenues for improvement of the TSFZ method. These results demonstrate that the TSFZ technique can produce h-BN crystals of high crystallinity and high purity while simultaneously providing a route to significantly increased volume.

\section{Crystal growth methodology}

\begin{figure}
  \centering
    \includegraphics[width=8.6cm]{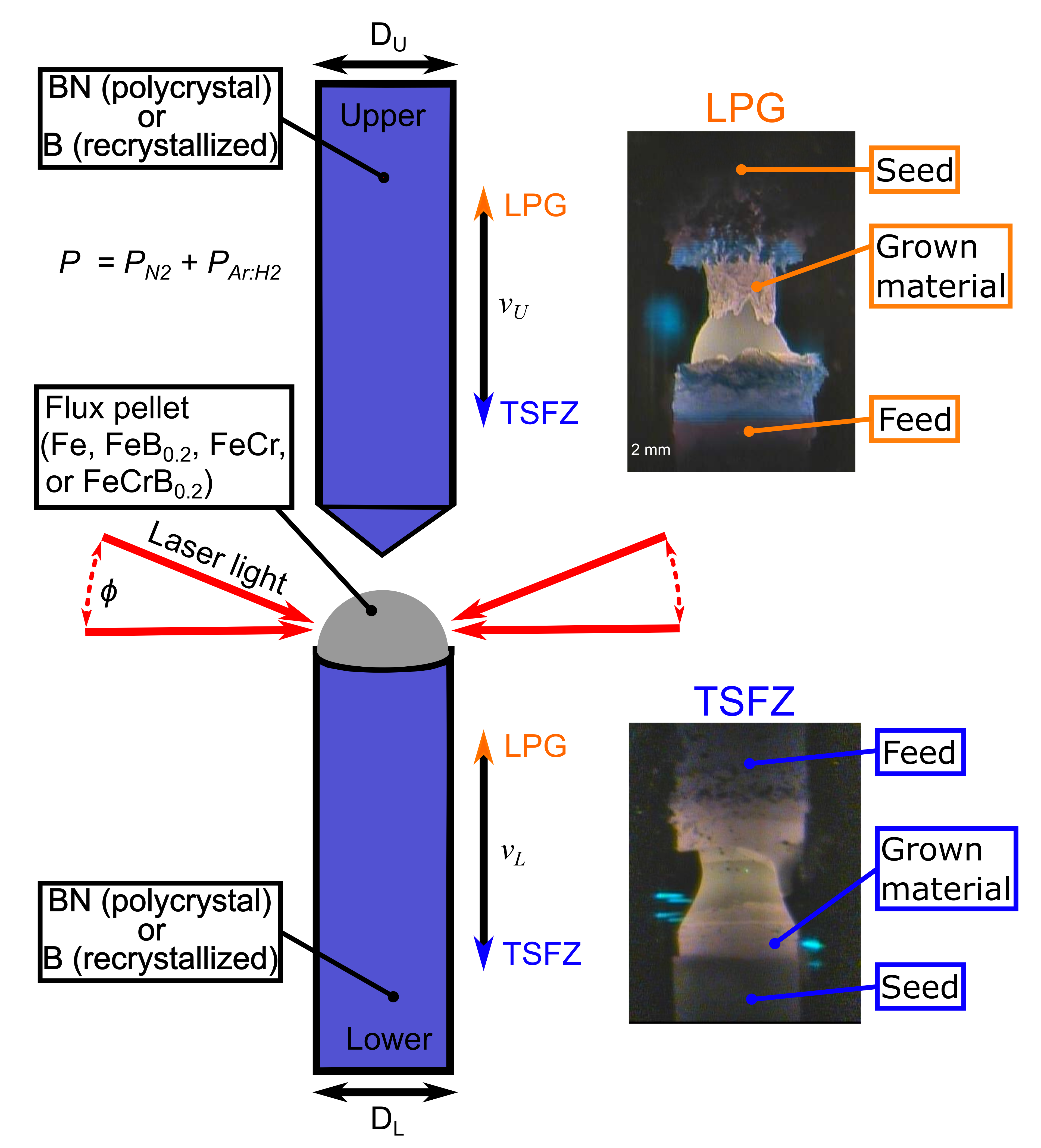}
    \caption{\textrm{(Left) Illustration of the growth geometries (TSFZ and LPG) used in this work. \textit{v}$_{U}$ and \textit{v}$_{L}$ indicate the direction of motion of the top and bottom rods, respectively, while $D_{U}$ and$D_{L}$ indicate the rod diameters. $\phi$ is the adjustable angle of the lasers relative to the translation axis in the PARADIM furnace; in the UCSB furnace $\phi$ = 0\textdegree. $P$, $P_{N2}$, and $P_{Ar:H2}$ are the total pressure and partial pressures of N$_{2}$ and the Ar:H$_{2}$ mix, respectively. (Right) Photographs of growths in-process using either LPG (top, sample \textbf{10}) or TSFZ (bottom, sample \textbf{4}).} \label{fig1}}
\end{figure}

A majority of the TSFZ growths in this work were conducted using the tilting, laser-based floating-zone system at the Platform for the Accelerated Realization, Analysis and Discovery of Interface Materials (PARADIM) crystal growth facility (Crystal Systems, Inc., FD-FZ-5-200-VPO-PC), which utilizes 5 $\times$ 200 W lasers ($\lambda$ = 976 nm). Additional growths were conducted at the University of California, Santa Barbara (UCSB) using a custom-built, high-pressure, laser-based floating-zone system \cite{schmehr2019,gomez2024} which utilizes 7 $\times$ 200 W lasers ($\lambda$ = 1070 nm). Two different growth geometries were tested to find the optimum conditions for h-BN. In standard TSFZ, a pellet of the flux is balanced on top of a seed rod made of polycrystalline h-BN, with a feed rod installed above. To initiate the growth the flux is melted, fusing it to the seed rod, and the feed rod is translated downwards until it connects to the molten flux and forms a stable molten zone. By translating the feed and seed rods downwards in tandem the feed rod dissolves into the flux and h-BN recrystallizes onto the seed, causing the molten flux to "travel". An alternative geometry – similar to Czochralski growth – is to place the feed on the bottom, melt the flux in the same manner as in TSFZ, and then dip a seed rod into the molten flux from above. The seed rod is then translated upwards to recrystallize h-BN from the zone while the feed rod may or may not be translated depending on its size relative to the seed. We refer to this geometry as "laser pedestal growth" (LPG) \cite{feigelson1985,andreeta2010} to distinguish it from the standard geometry, which is referred to as TSFZ. To summarize, in the TSFZ technique the feed rate (\textit{v}$_{feed}$) corresponds to \textit{v}$_{U}$ in Fig. \ref{fig1}, while the crystal growth rate (\textit{v}$_{grow}$) corresponds to \textit{v}$_{L}$. In LPG the situation is reversed.

The main parameters relevant to these geometries are illustrated in Fig. \ref{fig1} and include the rod diameter ($D$), rod movement rate and direction (\textit{v}), rod composition, flux composition, and total gas pressure ($P$). In Table S1 in the Supplemental Information (SI) we show a summary of the growth parameters for all samples considered in this work, which are referred to by their sample numbers (e.g. \textbf{1}, \textbf {2}, \textbf{3}) in the following.

\subsection{Flux preparation}

Two fluxes with different transition metal stoichiometries were tested: Fe and FeCr. Additional variants of these fluxes were tested by adding B prior to the growth (presaturation) to form nominal compositions of FeB$_{0.2}$ and FeCrB$_{0.2}$.  The fluxes were created by arc melting stoichiometric amounts of Fe (Kurt J. Lesker Co., 99.95\%), Cr (Kurt J. Lesker Co., 99.95\%), and amorphous B (Thermo Scientific, 98\%) together under an Ar atmosphere. The Fe:B ratio was chosen based on the solubility of B in liquid Fe \cite{kubaschewski2013}. The resulting ingots were then sectioned into smaller amounts and arc melted again on a flat hearth to create flat-bottomed, dome-shaped pellets which were more easily balanced on the seed rod. The mass of the flux pellet for a growth was chosen based on the diameter of the seed rod: typically $\approx$ 0.2 g for a 3 mm diameter and $\approx$ 0.5 g for a 5 --- 6 mm diameter. These masses yielded a sufficient volume of molten flux such that the full surface of the rods could be fused to the flux.

\subsection{Gas environment}

In the growths conducted at PARADIM, concurrent flows of N$_{2}$ and Ar:H$_{2}$ = 95:5 (by volume) were utilized. The N$_{2}$ was supplied either from a standard gas cylinder (Airgas, 99.999\%) or a high-pressure liquid dewar (Airgas, 99.999\%). In later growths, the N$_{2}$ was passed through a room temperature getter (Nupure, Eliminator Model 200 CA). Growths were conducted at a total pressure $P$ = 1, 5 or 7 bar using a N$_{2}$ flow rate of 0.5 L/min and an Ar:H$_{2}$ flow rate of 0.05 L/min, yielding a N$_{2}$ partial pressure of $P_{N2} \approx 0.9P$. Note that in the following we always refer to the total pressure used in a particular growth. The chamber was purged at these flow rates for 15 minutes prior to setting the final pressure. For the growths at UCSB a static pressure of pure N$_{2}$ was utilized because the Inconel pressure chamber is not compatible with H$_{2}$. Growths were conducted at $P = P_{N2}$ = 50 bar with the N$_{2}$ supplied from a standard gas cylinder (Airgas, 99.999\%) and flowed through a room temperature getter (Nupure, Eliminator Model 1000 CG). The chamber was initially purged at atmospheric pressure using a 1 L/min flow of N$_{2}$ for 15 minutes; the pressure was then cycled several times between 30 and 3 bar before setting the final pressure. 

\subsection{Rod composition, size, and movement rate}

When h-BN was used as the feed rod, commercially available hot-pressed, polycrystalline samples were utilized. For the growths at PARADIM the rods (Momentive, HBN-type) contained roughly 5 weight \% B$_{2}$O$_{3}$, as received. Prior to growth, these polycrystalline h-BN rods were annealed at 1400 \textdegree C under vacuum (5 $\times$ 10$^{-5}$ torr) for 24 hours, resulting in significant evaporation of B$_{2}$O$_{3}$. SEM/EDS analysis of these rods post degassing showed no sign of any elements besides B and N. TSFZ growths were performed with both 3 mm and 6 mm diameter polycrystalline h-BN feed and seed rods. In some cases, a pure B crystal (diameter 2 -- 5 mm) was utilized as the feed rod (see below). For the LPG geometry growths, the feed was either a 5 mm polycrystalline h-BN rod or a 5 -- 6 mm B crystal. These growths were seeded using a boule from a previous TSFZ growth (either the sample shown in Fig. \ref{fig2}b or that in Fig. S1a) which was kept attached to its polycrystalline seed rod.  For the TSFZ growths at UCSB, 3 mm diameter polycrystalline h-BN rods (Premier Ceramics) were used as the feed and seed. These rods were quoted by the supplier as “binder free” and were not degassed before use. For the UCSB and PARADIM TSFZ growths with polycrystalline h-BN rods the tip of the feed rod was sanded into a thin conical shape prior to use. Additionally, a small divot was drilled into the polycrystalline seed rod to secure the flux prior to melting. 

Pure B crystals for use as feed rods in h-BN growths were produced using the standard floating-zone method. Amorphous B powder (Thermo Scientific, 98\%) was pressed into rods using a clean rubber balloon (3 or 5 mm diameter) and a cold isostatic press pressurized to 3450 bar. The pressed B rods were sintered under a low Ar flow at 1400 \textdegree C for 24 hours prior to growth. The B crystals were grown under a 1 bar Ar atmosphere (Airgas, 99.999\%) with a flow rate of 0.5 L/min. The crystal growth rate was 7.5 mm/h and a variable feed rate of 5 – 10 mm/h was used to control the crystal diameter. The lasers were tilted at $\phi$ = 4\textdegree, which helped to stabilize the growth.

In all PARADIM TSFZ h-BN growths, \textit{v}$_{grow}$ was set to 0.1 mm/h and \textit{v}$_{feed}$ was set to 0.1 – 0.4 mm/h depending on the ratio between the feed and seed rod diameters and the composition of the feed rod. In LPG growths, \textit{v}$_{grow}$ was set to 0.1 mm/h. The feed rod was not moved during LPG growth because the diameter of the seed rod was substantially smaller than that of the feed. For the growths at UCSB it was not possible to use the same \textit{v}$_{grow}$ = 0.1 mm/h because the slowest translation speed achievable with the high-pressure compatible drives is 0.4 mm/h. As a result, the high-pressure growth (sample \textbf{3}) utilized \textit{v}$_{grow}$ = \textit{v}$_{feed}$ = 0.4 mm/h .

In the TSFZ growths at PARADIM  and UCSB, the seed rotation rate was 8 rpm and the feed rotation rate was typically 9 rpm. Over the course of growths with a h-BN feed the interface between the feed rod and flux occasionally deteriorated due to uneven dissolution of the feed. A faster feed rotation rate of 15 rpm was then briefly employed ($\approx$ 1 h) until the interface quality improved. For the LPG growths at PARADIM the rotation rates were as follows: sample \textbf{6} used a feed rotation rate of 6 rpm and a seed rotation rate of 8 rpm. Sample \textbf{10} used a feed rotation of 4 rpm and a seed rotation rate of 12 rpm.

\subsection{Flux temperature}

The temperature in these growths was controlled by changing the applied power of the laser system. The steady-state power was set relative to the power level needed to completely melt the flux pellet, which varied directly with the mass. In initial experiments using a FeCr flux and a h-BN feed rod, the laser power during steady-state growth was set substantially higher than the initial laser power needed to melt the flux (typically an $\approx$ 50\% increase). In later growths, the laser power during steady state growth was set to a lower value, typically $\approx$ 5 – 10\% over the initial melting power. This lower laser power was preferable, as it was still sufficient for h-BN dissolution by the flux but led to less volatility. This volatility occurs regardless of flux composition and most likely stems from the finite vapor pressure of Fe and Cr since SEM/EDS measurements (not shown) indicate that it contains the metals utilized in the flux. Anecdotally, the rate of volatilization was observed to be greater (for equal laser power) in the presence of Cr, consistent with the higher vapor pressure of Cr. The ability to tilt the laser diodes in the PARADIM furnace allows for adjustment of the temperature gradient across the interface between the molten flux and the upper rod. A small inclination ($\phi$ = 2\textdegree) enhanced dissolution of the feed rod in the TSFZ geometry and was therefore employed in TSFZ growths with h-BN feed rods. In the case of B feed rods, no inclination was used ($\phi$ = 0\textdegree) to avoid direct heating of the B crystal and subsequent thermally induced cracking. All LPG growths were conducted with no inclination ($\phi$ = 0\textdegree).

\section{Characterization methodology}

\subsection{Micro computed X-ray tomography}

Micro computed X-ray tomography (microCT) data were collected using a Bruker SkyScan 1172 operating at 100 kV and 100 $\mu$A with downstream Al and Cu filters in place. MicroCT allows for visualization of the internal structure of the grown samples, providing a straightforward means of evaluating flux incorporation and crystal morphology to illuminate the effects of different growth parameters. The pixel size ranged from 2 – 6 $\mu$m, depending on the highest level of magnification which kept the sample in the field of view. Data was collected by rotating the sample 360\textdegree$ $ and collecting exposures every 0.25\textdegree$ $– 0.35\textdegree. Reconstruction of the two-dimensional exposures into a full three-dimensional data set was done using the NRecon software \cite{Nrecon} with ring artifact removal and beam hardening correction algorithms applied. The reconstructed data consists of a collection of voxels, each containing an absorption value. This data was then visualized using the Dragonfly software, Version 2021.1 for Windows (Object Research Systems Inc, Montreal, Canada, 2020; software available at http://www.theobjects.com/dragonfly). 

In Dragonfly, the absorption value of each voxel is mapped to a grey scale value with, by default, the minimum and the maximum absorption values mapped to the minimum and maximum greyscale values, respectively. The substantial difference in absorption between h-BN and the metal fluxes leads to excellent contrast within the volume of the sample. However, the weak absorption of X-rays by h-BN leads to poor contrast with the empty regions of the sample chamber. To enhance this contrast, the mapping was modified such that an absorption value less than the maximum was mapped to the maximum greyscale value; this effectively condenses the color scale, making it so all voxels above the chosen value  correspond to the maximum grey scale value in the image.

\subsection{Raman spectroscopy}

Raman spectroscopy data was collected in order to evaluate the crystallinity of grown crystals via the metric of the full-width-at-half-max (FWHM) of the measured phonon modes. All measurements were conducted at room temperature ($T$ = 295 K) in the backscattering geometry. Most measurements were conducted using a Horiba LabRAM HR Evolution Raman microscope system with a $\lambda$ = 532 nm laser. The laser light was focused by a 100$\times$ objective lens to a spot size of $\approx$ 1 $\mu$m and the laser power was kept below 0.6 mW to avoid laser-induced heating. The scattered light was dispersed by an 1800 grooves/mm grating and detected by a thermoelectrically cooled CCD. Additional measurements (Fig. S9) were conducted using a Horiba JY T64000 spectrometer equipped with an Olympus microscope and a $\lambda$ = 514.6 nm laser. The light was focused to a spot size of $\approx$ 2 $\mu$m and the laser power was kept below 1 mW. In some cases samples were cut and polished prior to measurement, while in other cases the samples were measured as-synthesized; this distinction is noted in the relevant figure captions.

\subsection{Photoluminescence spectroscopy}

Photoluminescence (PL) spectroscopy in the deep UV range was performed at low temperature ($T$ = 8 K) with the samples held on the cold finger of a closed-cycle cryostat. In addition to the intrinsic PL signal, various extrinsic (i.e. disorder-driven) PL processes are known to occur in h-BN. Evaluation of the corresponding PL signals provides a more detailed probe of crystal quality. The PL was excited above the bandgap by the fourth harmonic of a CW mode-locked Ti:sapphire oscillator with 140 fs-pulses at 80 MHz repetition rate. The detection system was composed of a $f$ = 500 mm Czerny–Turner monochromator, equipped with an 1800 grooves/mm grating blazed at 250 nm, and a CCD camera (Andor Newton 920). The spectra were recorded across the energy range 3.8 -- 6 eV with an excitation energy of 6.2 eV ($\lambda$ = 198 nm) and an incident power of around 40 $\mu$W. Unless stated otherwise, all samples were cut and polished before measurement.

\subsection{X-ray diffraction}

X-ray diffraction (XRD) data were collected at room temperature ($T$ = 293 K) using a Bruker D8 Focus diffractometer employing Cu K$\alpha$ radiation ($\lambda$ = 1.5406\r{A}) and a position sensitive Lynx Eye detector. A Ni filter was employed to minimize the amount of Cu K$\beta$ radiation in the incident beam. The collimation consisted of a 0.6\textdegree $ $ Soller slit on the incident beam and an 8 mm anti-scatter slit and 2.5\textdegree $ $ Soller slit on the diffracted beam. The measured samples were either ground powders adhered to glass slides with grease, or polished sections cut from as-synthesized boules mounted with clay. Scans were measured across 2$\Theta$ = 5\textdegree$ $– 60\textdegree$ $ with a step size of 0.0186\textdegree/step and a count time of 0.35 – 1.13 s/step. 

\subsection{Scanning electron microscopy}

Scanning electron microscope (SEM) backscattered-electron images were collected using a JEOL JSM-IT100 SEM, with a 20 kV acceleration voltage. Samples (either cut and polished or unmodified) were mounted in the SEM using carbon tape on an Al stud. The crystal composition was measured by energy dispersive X-ray spectroscopy (EDS) – with the same acceleration voltage – to collect both spot scans and two-dimensional maps.

\subsection{Magnetometry}

Taking advantage of the large difference in the magnetic response between h-BN and the various flux compositions, magnetization versus field data were collected as an indirect means of further investigating the nature of flux incorporation. Data were taken at $T$ = 300 K across the field range $\mu_{0}H$ = \textpm 7 T using a superconducting quantum interference device (SQUID) magnetometer (Quantum Design, MPMS3) operating in the DC mode. Samples were mounted either by wrapping in thin polypropylene foil and securing inside a plastic straw ("chunk" sample) or by mounting to a quartz paddle with GE varnish ("flakes" sample). 

\section{Results}

\begin{figure*}
  \centering
    \includegraphics[width=16.4cm]{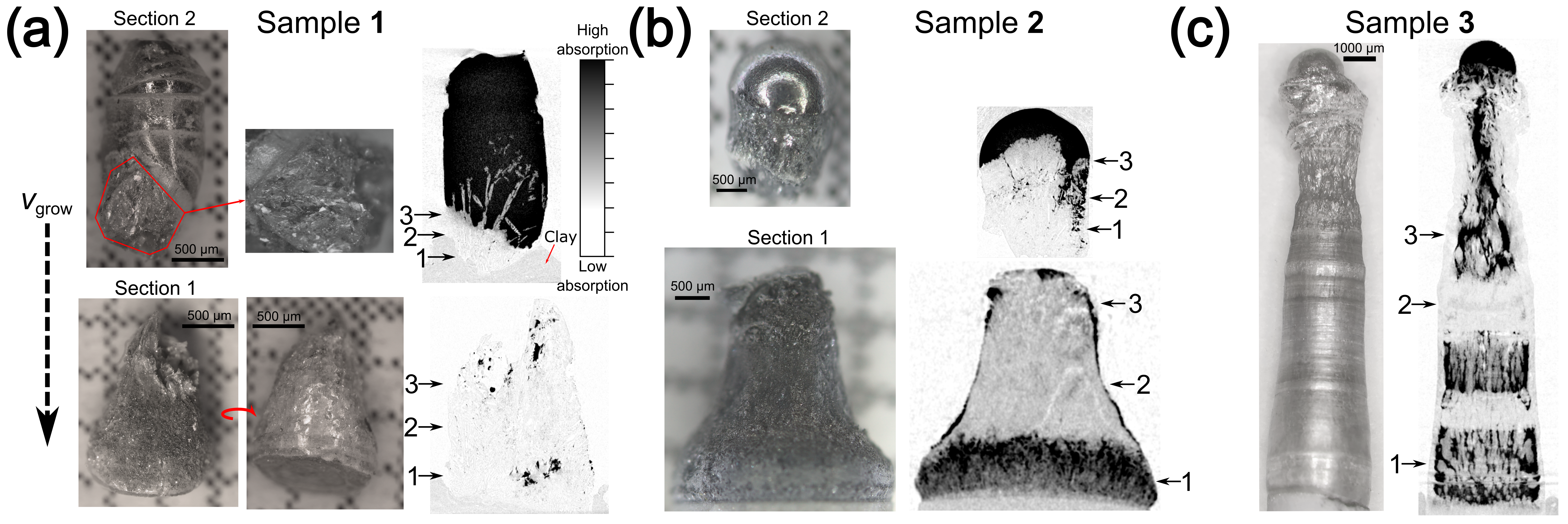}
    \caption{\textrm{Optical microscopy images (left) and corresponding microCT data (right) for three representative samples produced in this work. The growth conditions for each sample are provided in the caption. The dashed black arrow on the left shows the growth direction for all samples. The solid black arrows on the microCT images indicate the location of the orthogonal slices shown in Fig. \ref{fig3}, with higher numbers corresponding to regions from later in the growth. The color scale is schematic and applies to all of the presented microCT data.  (a) Fe flux, $P$ = 5 bar, \textit{v}$_{grow}$ = 0.1 mm/h, h-BN feed, TSFZ, $\phi$ = 2\textdegree, $D_{U}$  = $D_{L}$ = 3 mm.  (b) FeCrB$_{0.2}$ flux, $P$ = 5 bar, \textit{v}$_{grow}$ = 0.1 mm/h, h-BN feed, TSFZ, $\phi$ = 2\textdegree, $D_{U}$  = $D_{L}$ = 3 mm.  (c) FeCrB$_{0.2}$ flux, $P$ = 50 bar, \textit{v}$_{grow}$ = 0.4 mm/h, h-BN feed, TSFZ, $\phi$ = 0\textdegree, $D_{U}$  = $D_{L}$ = 3 mm. For (a) and (b), the sample broke into two sections upon removal from the furnace. In both cases section 1 is from earlier in the growth and section 2 is from later in the growth.} \label{fig2}
}
\end{figure*}

\begin{figure}
  \centering
    \includegraphics[width=8.6cm]{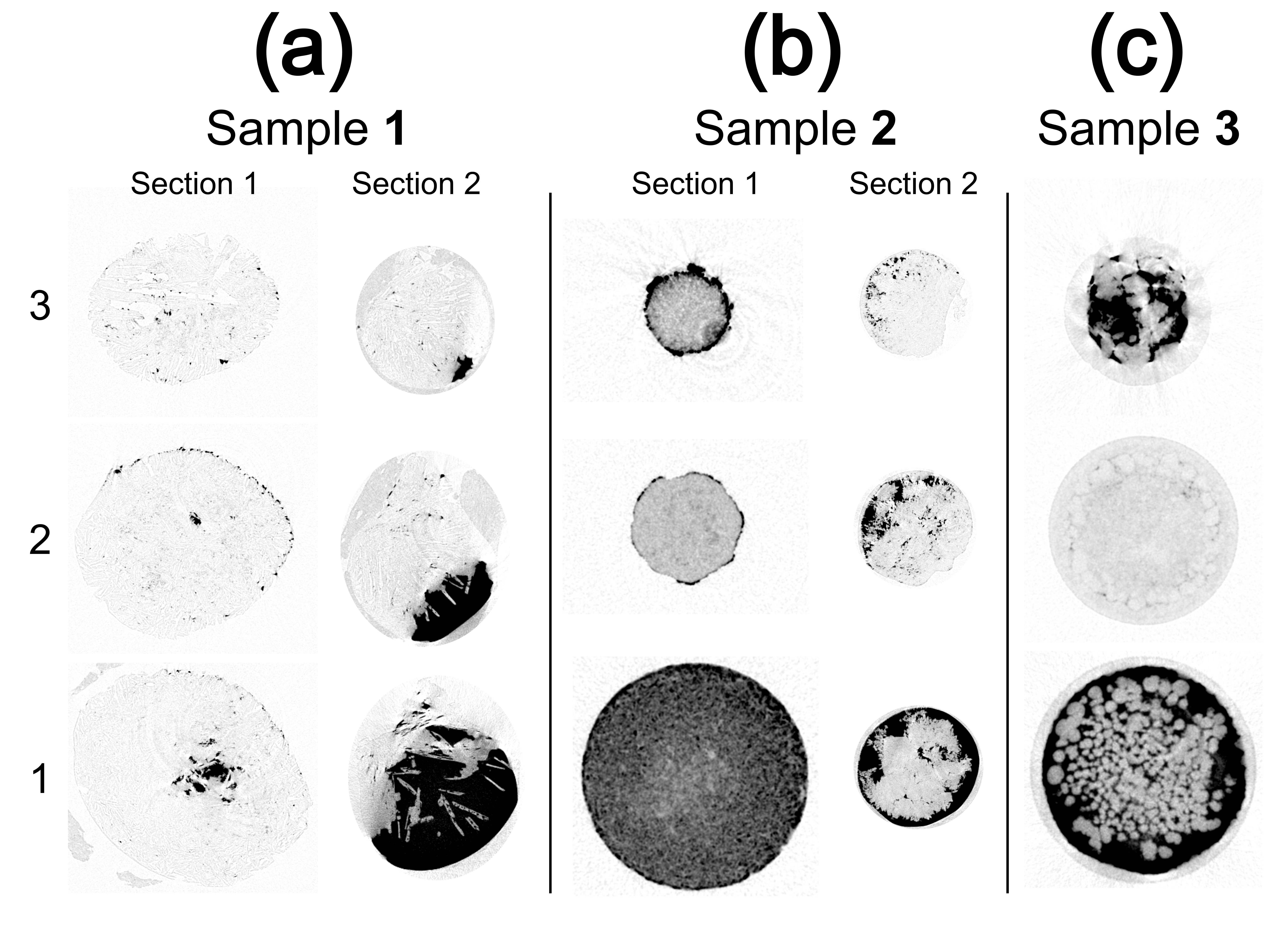}
    \caption{\textrm{Corresponding microCT cross-sectional images for the three TSFZ-grown samples shown in Fig. \ref{fig2}. The number labeling each row of images corresponds to the same numbered black arrow in the corresponding panel of Fig. \ref{fig2}}. \label{fig3}
}
\end{figure}

\begin{figure}
  \centering
    \includegraphics[width=8.6cm]{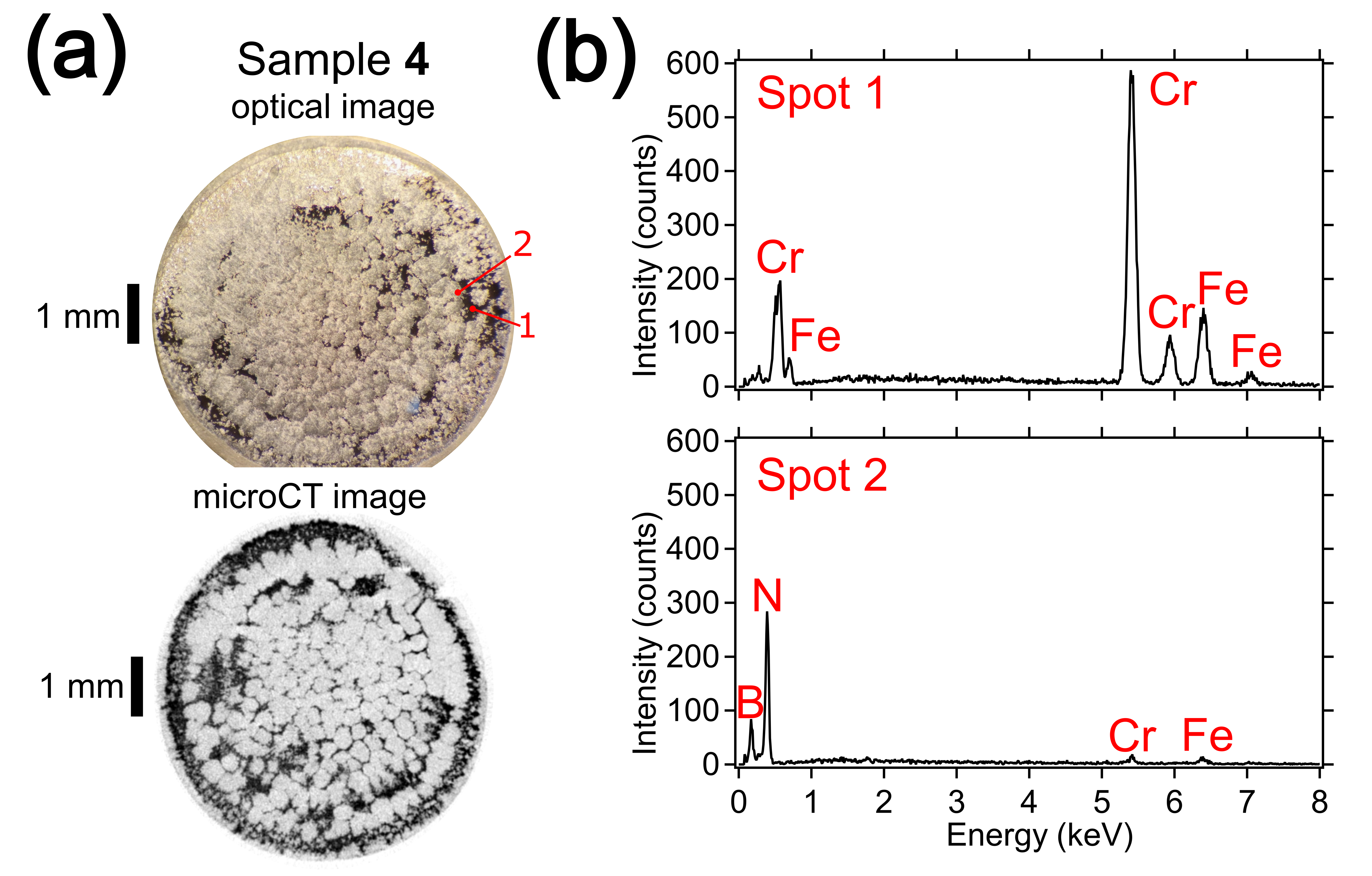}
    \caption{\textrm{(a) Optical image of a cut and polished section from sample \textbf{4}, grown by TSFZ using an FeCr flux, $P$ = 1 bar, \textit{v}$_{grow}$ = 0.1 mm/h, h-BN feed, $\phi$ = 0\textdegree, and $D_{U}$  = $D_{L}$ = 6 mm. The microCT image corresponds to a region near to the surface shown in the optical image, though from slightly deeper into the sample volume. (b) EDS spectra collected at the labeled positions in (a). Spot 1 is on the darker matrix and spot 2 is on the colorless crystalline material.}\label{fig4}
}
\end{figure}

\begin{figure*}
  \centering
    \includegraphics[width=16.4cm]{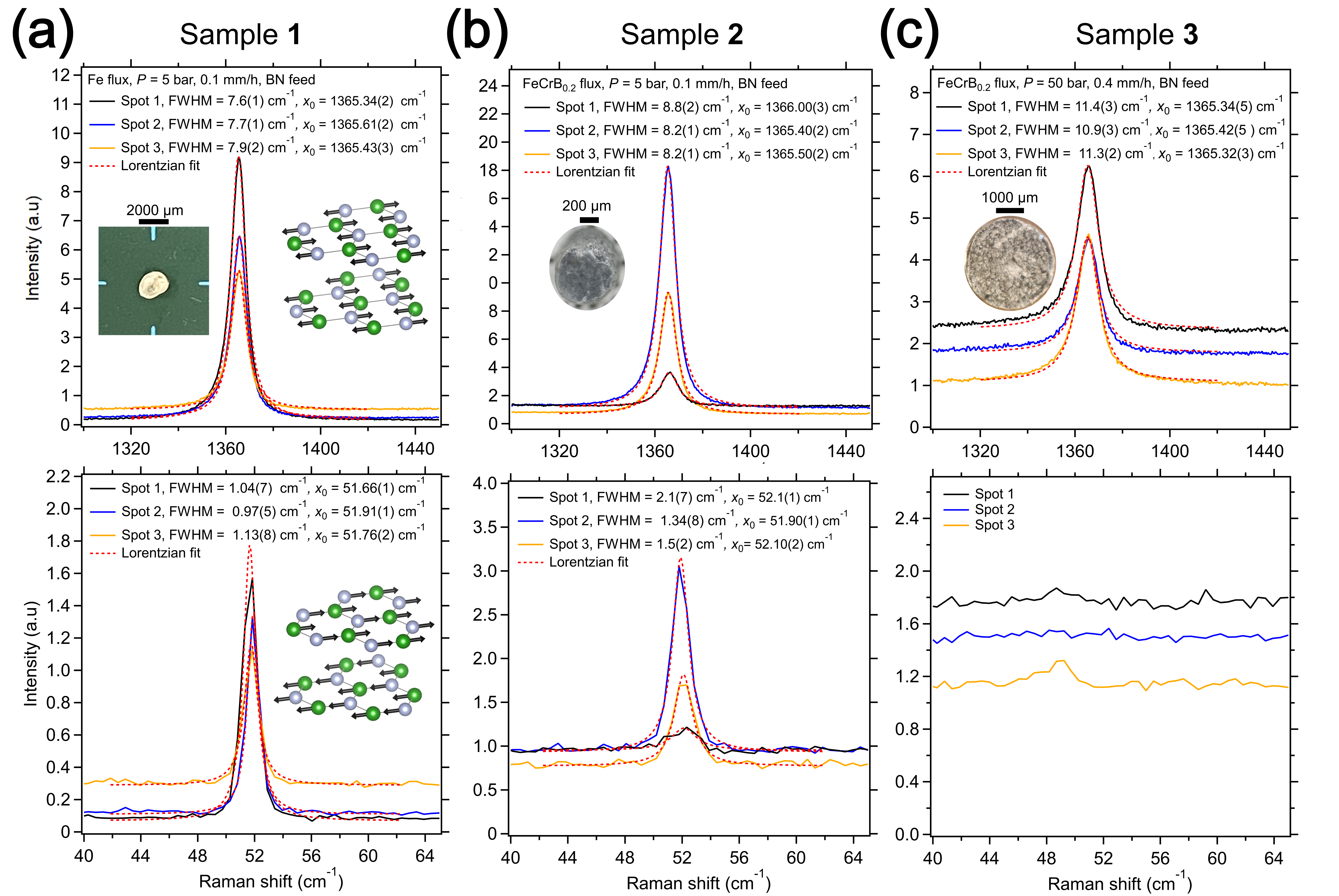}
    \caption{\textrm{Raman data for three representative TSFZ-grown h-BN samples showing the E$_{\mathrm{2g}}$ intralayer (top) and interlayer (bottom) modes. The dashed red line is a fit with a Lorentzian peak on a contstant plus linear background; $x_{0}$ is the fitted position of the Lorentzian. The labels above each set of panels indicates the measured sample (inset optical image) which are those shown in Fig. \ref{fig2}. (a) Intralayer averages: FWHM = 7.7(2) cm$^{-1}$, $x_{0}$ = 1365.46(4) cm$^{-1}$. Interlayer averages: FWHM = 1.1(1) cm$^{-1}$, $x_{0}$ = 51.78(9) cm$^{-1}$. The measured piece was extracted from section 2 of this sample (area enclosed in red line in Fig. \ref{fig2}a) following soaking in aqua regia and washing with ethanol; no cutting and polishing was done. The ball-and-stick insets show the atomic motions corresponding to each phonon mode. (b) Intralayer averages: FWHM = 8.4(2) cm$^{-1}$, $x_{0}$ = 1365.63(4) cm$^{-1}$. Interlayer averages: FWHM = 1.4(2) cm$^{-1}$, $x_{0}$ = 52.00(2) cm$^{-1}$. The measured piece is from section  1 of this sample and was cut and polished before measurement. Due to the low intensity, spot 1 was excluded from the interlayer averages. (c) Intralayer averages: FWHM = 11.2(5) cm$^{-1}$, $x_{0}$ = 1365.36(8) cm$^{-1}$. Interlayer averages: N/A. The measured piece was cut and polished before measurement.} \label{fig5}
}
\end{figure*}

\begin{figure}
  \centering
    \includegraphics[width=8.6cm]{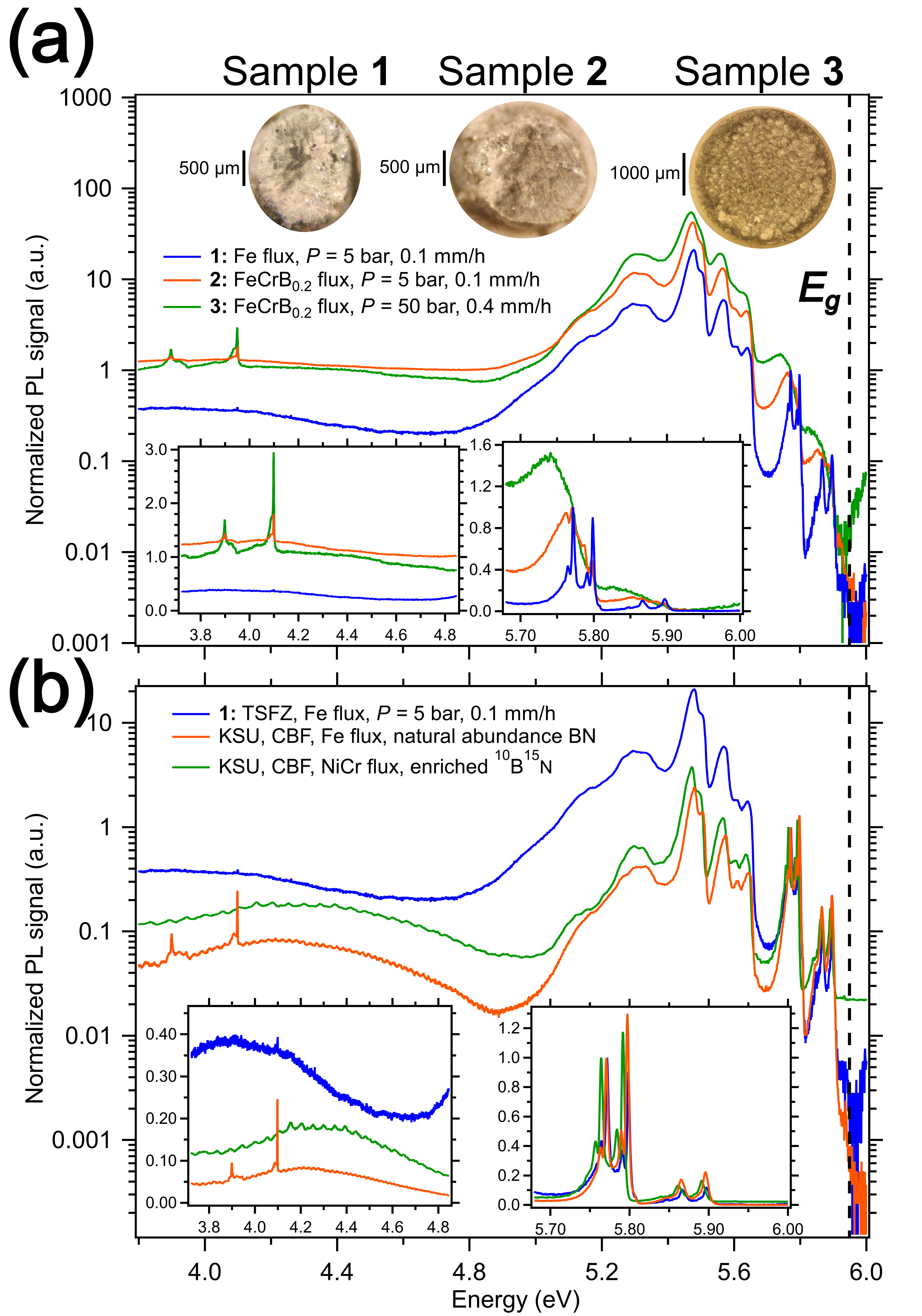}
    \caption{\textrm{(a) PL spectra collected on three TSFZ-grown samples shown in Fig. \ref{fig2} and Fig. \ref{fig3}, following cutting and polishing. The numbers in the legend correspond to the sample numbers. The data are normalized to the value of the signal at $E$ = 5.77 eV. The dashed black line shows the energy of the indirect gap ($E_{g}$ = 5.955 eV \cite{cassabois2016}). Optical images of the measured cut and polished surfaces are shown in the corresponding insets. (b) Comparison of the PL spectrum of the highest-quality TSFZ-grown sample (\textbf{1}) to two different samples grown by the CBF method at Kansas State University.} \label{fig6}
}
\end{figure}

In Fig. \ref{fig2} we show optical images of samples (\textbf{1}, \textbf{2} and \textbf{3}) grown under different conditions side-by-side with corresponding microCT data. Optical images and corresponding microCT data of other exemplar samples are shown in Fig. S1 and Fig. S2, respectively. The range in morphology and outward appearance of the samples is readily apparent. Different regions of individual samples also varied, such as in the sample shown in Fig \ref{fig2}c, where the surface morphology shows a clear crossover between the upper and lower regions. However, most of the samples show at least some colorless/transparent regions on their outer surface, and all show such regions internally (see optical images shown in Fig. \ref{fig4}, \ref{fig5} and \ref{fig6}). MicroCT provides an efficient first-pass characterization of these samples, yielding a three-dimensional view of the sample volume that provides insight into the nature and morphology of the crystallized material. The microCT images of the three samples shown in Fig. \ref{fig2} represent a slice through the center of the samples along the growth direction. In Fig. \ref{fig3} we show cross-sectional slices in the direction orthogonal to the growth direction, with each slice taken at the corresponding black arrow in Fig. \ref{fig2}. Distinct regions of high and low absorption were present in all samples.

EDS analysis provides elemental information regarding the distinct regions seen in the microCT data. Fig. \ref{fig4}a shows an optical image and corresponding microCT cross-section for a sample grown from an FeCr flux (sample \textbf{4}, not shown in Fig. \ref{fig2} or \ref{fig3}). The colorless crystalline regions of the optical image correspond to the low absorption regions seen in the microCT cross-section. The darker matrix surrounding these colorless regions corresponds to the high absorption regions. An SEM back-scattered electron image of the sample is shown in the SI, in Fig. S3a. EDS spot scans taken within the two regions (Fig. \ref{fig4}b) show distinct compositions: the spectra in the darker matrix region (spot 1) is dominated by emission lines from Fe and Cr, while the spectra in the transparent crystalline region shows clear B and N emission lines, with very minor Fe and Cr contributions. Due to their low atomic numbers we do not attempt a quantitative analysis of the B and N stoichiometry by EDS. 

The previous data demonstrate clearly that the transparent colorless regions are a B and N containing phase. Powder XRD on crushed samples (see Fig. S4a,b) confirms the formation of the hexagonal BN structure under both $P$ = 7 bar and $P$ = 50 bar. Additional XRD data collected on cut and polished surfaces show very strong (00$L$)-type preferred orientation for some regions of certain samples (Fig. S4c,d), such that other peaks are almost totally absent. These regions are identifiable in the microCT data by the occurrence of circular, columnar low absorption regions (see, for example, the region marked "1" in Fig. \ref{fig2}c and the corresponding cross-section in Fig. \ref{fig3}c) which appear to correspond to be distinct h-BN grains. However, other grain morphologies are present, such as the more undifferentiated agglomerations seen in Fig. \ref{fig2}a,b and Fig. \ref{fig3}a,b. XRD data (not shown) collected on other cut and polished sections of the samples in Fig. \ref{fig2} (shown in the insets in Fig. \ref{fig6}) show all of the expected h-BN Bragg peaks and significantly less (00$L$)-type preferred orientation. Under certain growth conditions (Fe flux, $P$ = 7 bar, B feed, TSFZ) a unique morphology was observed consisting of majority polycrystalline material, with colorless, dendritic, crystalline flakes forming in the core of the sample (see Fig. S1c and S1d).

The microCT and SEM data also demonstrate the clear occurrence of macroscopic regions of trapped flux (segregated from the h-BN regions) within the grown samples. The volume fraction of trapped flux is quantifiable via segmentation of the microCT data based on absorption values, as shown and described in Fig. S6. This analysis yields the following flux volume fractions for the samples shown in Fig. \ref{fig2}: sample \textbf{1} = 2 volume \%, sample \textbf{2} = 38 volume \%, and sample \textbf{3} = 14 volume \%. In the case of samples grown at $P$ = 1 – 7 bar with $v_{grow}$ = 0.1 mm/h,  flux residue is often observed on the exterior of the sample as a surface layer (grey regions seen in Fig. \ref{fig2}a,b as well as Fig. S1a), as well as internally as inclusions. The amount of exterior flux tends to be reduced at the higher end of this pressure regime. For samples grown at $P$ = 50 bar with $v_{grow}$ = 0.4 mm/h, the trapped flux appears exclusively in the internal volume of the sample, with the exterior surface totally covered in h-BN. This behavior is shown in Fig. \ref{fig2}c , and was also observed in other samples (not shown) grown under the same conditions. As a comparison to the bulk samples, in Fig. S5 we show SEM/EDS data (as well as other characterization) on a transparent dendritic flake extracted from a sample grown with an FeB$_{0.2}$ flux. EDS spot scans from different regions of this sample show strong B and N emission lines, with no sign of Fe emission, indicating that minimal Fe enters the h-BN lattice itself during TSFZ growth. Additional investigation of flux incorporation into the grown sample by magnetization measurements is shown in the SI (Fig. S7).

We now turn to spectroscopic characterization of select samples. Raman spectra for three samples grown under different conditions are shown in Fig. \ref{fig5}. Comparison of samples \textbf{1} and \textbf{2} illustrates the effect of changing flux composition (Fe or FeCrB$_{0.2}$, respectively). Both samples show the intralayer (high-energy, E$_{\mathrm{2g}}$) and interlayer (low-energy, E$_{\mathrm{2g}}$) Raman modes characteristic of h-BN \cite{geick1966}, further confirming the crystalline structure. However, while there is no significant difference in peak positions, the FWHM for both modes is narrower in sample \textbf{1} than in sample \textbf{2}. Sample \textbf{3} shows the broadest intralayer mode, and the interlayer mode is not observed. Additional Raman data exploring the impact of flux inclusions on the Raman spectra are shown in Fig. S8. These data indicate that the Raman spectra of the h-BN grains are not strongly modified near the interface with the trapped flux.

In Fig. \ref{fig6} we show representative normalized PL data collected on pieces from the same three samples as for the Raman data (Fig. \ref{fig5}). The data in the energy range $E \approx$ 4.8 – 5.7 eV are qualitatively similar for all three samples, showing the same features. These features in the PL spectrum of h-BN have been assigned to bound exciton luminescence due to stacking faults and donor-acceptor pair (DAP) recombination driven by charged defects \cite{museur2008a,museur2008b}. In contrast, distinct differences are observed at higher ($E >$ 5.7 eV) and lower ($E <$ 4.8 eV) energies. At lower energy (left lower inset, Fig. \ref{fig6}a), samples \textbf{2} and \textbf{3} show two clear peaks centered at $E \approx$ 3.9 eV and 4.1 eV, which have been associated with a carbon-related defect \cite{bourrellier2016,museur2008a,museur2008b,moore1972,katzir1975}, most likely a C$_{2}$ dimer \cite{plo2024}. Sample \textbf{1} shows a substantial reduction in these peaks, with no observable intensity at 3.9 eV. A very slight peak is resolvable at 4.1 eV the height of which, relative to the background on the high energy side, is approximately 14 times smaller than the same feature in sample \textbf{2} and 51 times smaller than for sample \textbf{3}. We also see substantial differences in the higher energy ($E >$ 5.7 eV) region. Sample \textbf{1} clearly shows the sharp peaks associated with phonon-assisted free exciton luminescence \cite{cassabois2016}.  Sample \textbf{2} shows resolvable remnants of these features although they are substantially broadened. The signal corresponding signal in sample \textbf{3} shows no resolvable peaks. 

\section{Discussion}

From the previous results the most apparent growth variable affecting the h-BN crystal quality is the flux composition. The published CBF work shows that, for that technique, the choice of Fe or Fe-Cr yields relatively minor differences in crystal quality and has no significant impact on crystal size \cite{liu2017,li2020a,li2021}. This is unexpected since the solubility of N in Fe alone is significantly lower than in Cr alone \cite{raghavan1987,deluca1973} and the increased concentration of Cr is well known to increase N solubility in Fe-Cr alloys \cite{frisk1990}. Consistent with the CBF results, the microCT data indicates that, with other parameters held constant at appropriately optimized values, both Fe (sample \textbf{1}) and FeCr (samples \textbf{2} and \textbf{5}) can yield relatively large regions of largely flux-free h-BN via TSFZ. 

However, in contrast to the CBF results with metal fluxes, optical characterization indicates that  higher quality is obtained from TSFZ growth without the use of Cr in the flux. In the PL data (Fig. \ref{fig6}a) the sharp reduction in the signals near $E$ = 3.9 eV and 4.1 eV for sample \textbf{1} (Fe flux) as compared to \textbf{2} (FeCrB$_{0.2}$ flux) qualitatively indicates a higher purity. Furthermore, sample \textbf{1} shows the clearest phonon-assisted emission lines. Since samples \textbf{1} and \textbf{2} in Fig. \ref{fig6} were grown under otherwise identical conditions, we conclude that the difference in flux composition leads to the differences in the PL spectrum. The Raman spectroscopy results show a similar comparison, with the FWHM of the two modes consistently narrower for sample \textbf{1} than for sample \textbf{2}. This is further confirmed by the more limited Raman data set (Fig. S9) from a different section of sample \textbf{2}. This data also shows a broader FWHM for the intralayer mode than sample \textbf{1}, confirming that this behavior is consistent throught the sample.

Overall, these results indicate that including Cr in the flux, under otherwise identical growth conditions, worsens the crystallinity of samples grown by TSFZ. This may be related to the formation of Cr nitrides, which were observed via PXRD in samples with trapped flux (Fig. S4b). It seems possible that the Cr nitrides sequester a portion of the dissolved N in the flux, shifting the flux chemistry and leading to more disorder in the recrystallized h-BN. Again, these results are surprising since the optical properties of CBF samples grown with Fe and Fe-Cr are similar \cite{liu2017,li2020a,li2021} and inherent differences in the TSFZ and APHT CBF techniques are likely the cause. In the CBF technique the h-BN forms as a thin layer on the outside of the flux as it cools and the flux is not actively mixed. In contrast, material is continuously pulled from the flux in TSFZ and the flux undergoes significantly increased mixing. This difference between the techniques appears relevant since it likely influences the distribution of B and N in the flux, making them more uniform in TSFZ than in CBF growth. This may increase the formation rate of Cr nitrides and lead to an observable detrimental effect. Further experiments, characterization and modeling will be needed to fully understand these different behaviors. Such comparisons between the two techniques may prove informative about the details of the reaction pathways occurring within the flux.

Presaturation of the flux with B does not appear to be essential to synthesizing large, flux-free regions of h-BN. Repeating the growth shown in Fig. \ref{fig2}b but substituting FeCr for FeCrB$_{0.2}$ yielded a similar morphology (Fig. S1a) and proportion of h-BN by microCT (Fig. S2a). However, presaturation eliminates the need to dwell for long periods at the beginning of the growth in order to ensure that the molten flux has dissolved a suitable concentration of B. Presaturation also allows for tuning of the initial B concentration in the flux, which can then be maintained during the growth by setting appropriate translation rates for equalized B mass transfer. 

As noted previously, XRD on certain sample surfaces perpendicular to the growth direction (Fig. S4c,d) showed only (00$L$) peaks. This data gives information about the orientation of the growing h-BN grains and demonstrates that, in these particular regions of these samples, the h-BN grains are strongly oriented with the $c$-axis parallel to the growth direction. The corresponding columnar regions observed in the microCT data for such regions (see region marked "1" in Fig. \ref{fig2}c) indicate that this grain growth along the $c$-axis direction was sustained. The anisotropic layered structure of h-BN, with only weak van der Waals bonding between the individual planes, makes this observation surprising. Much faster growth is expected in the $ab$-plane, which is reflected in the thin, plate-like samples obtained with the CBF technique, where the extent of the crystal is always significantly larger in the $ab$-plane than along the $c$-axis. Repeatable continuous $c$-axis oriented growth via TSFZ would be beneficial for producing thicker h-BN crystals with an ideal form factor for exfoliation. However, we have only observed this growth morphology in two samples (samples \textbf{3} and \textbf{4}) which were synthesized with substantially different gas pressures and growth rates. As a result, we are unable to connect this growth morphology to a specific set of growth parameters and it is not clear if it simply occurs stochastically or if it is the result of an unidentified common aspect of these two growths. The XRD data for other polished surfaces of sample \textbf{3} and other growths (not shown) show a more varied grain orientation with respect to the growth direction, indicating a less uniform growth mode.

Based on previous CBF results using pure B with Ni-Cr and Fe-Cr fluxes \cite{liu2018,li2020a} at atmospheric pressure, it is clear that N$_{2}$ gas participates in h-BN crystallization. This is sensible since the partial pressure of N$_{2}$ gas controls the equilibrium concentration of N ($n_{N}$) in the metal flux according to Sieverts’ law ($n_{N} \sim \sqrt{P_{N2}}$) \cite{sieverts1929}.  Therefore a higher N$_{2}$ pressure can be expected to increase h-BN crystallization and/or improve crystal quality by preventing N vacancies. At lower pressures this appears, qualitatively, to be the case. Growths with $P$ = 1 bar did not reliably produce samples with as large h-BN regions as comparable growths at $P$ = 5 – 7 bar with the same flux compostion, feed rod composition and $v_{grow}$ 

The microCT data collected on TSFZ samples grown at $P$ = 50 bar  — including sample \textbf{3} in Fig. \ref{fig2}c and \ref{fig3}c, as well as others for which the data is not shown — indicate that relatively large regions of h-BN can be obtained under these growth conditions, although the issue of trapped flux remains. However, the Raman (Fig. \ref{fig5}c) and PL (Fig. \ref{fig6}a) data for sample \textbf{3} make it clear that the quality is significantly worse than samples \textbf{1} and \textbf{2}. Before discussing our interpretation of these data sets, it is important to note again that the $P$ = 50 bar growths were conducted at a significantly faster growth speed than the lower pressure growths ($v_{grow}$ = 0.4 mm/h and 0.1 mm/h, respectively), since 0.4 mm/h was the minimum obtainable speed on the high-pressure floating-zone furnace. As a result, we are unable to isolate which variable (higher $P$ or faster $v_{grow}$) is responsible for the reduced quality in sample \textbf{3}. Nonetheless, the extremely broadened features at high energy in the PL data for this sample indicate significant lattice disorder. This interpretation is consistent with the Raman data, in which the intralayer mode is substantially broadened compared to samples \textbf{1} and \textbf{3} and the interlayer mode is completely absent. The broadening of the intralayer mode is most likely driven by an increased defect density within the $ab$-plane, which is further supported by the increased intensity of the low-energy ($E <$ 4.8 eV) PL signals. The complete absence of the interlayer Raman mode indicates an extreme loss of coherence in the stacking order of the BN planes in this sample. 

While, based on our experiments alone, we are unable to provide a definitive conclusion about the source of the increased disorder in sample \textbf{3}, preliminary work by Sadovyi et al. on CBF growths with Ni and Ni-Cr fluxes at signficantly higher $P$ = 1000 bar may provide some indication \cite{sadovyi2024,sadovyi2024a}. The h-BN crystals produced at this pressure show very similar quality, as indicated by Raman spectroscopy, as crystals grown by CBF at $P$ = 1 bar from the same flux \cite{hoffman2014}. This indicates that, within the context of the CBF technique, increasing the pressure does not lead directly to higher quality. By analogy, this may be taken to imply that the lower quality of sample \textbf{3} is due to the faster $v_{grow}$ not the increased pressure. Sensitivity to the growth rate appears in-line with CBF results showing poorer quality with a faster cooling rate \cite{hoffman2014}, which is roughly analogous to the growth rate in TSFZ/LPG. However, given the inherent differences between the two techniques, especially the different levels of physical mixing of the flux and the difference in crystallization processes (directional for TSFZ/LPG and surface only for CBF), this comparison is not definitive regarding the impacts of pressure and $v_{growth}$. 

Regarding the utilized feed rod compositions, our results indicate that either are workable with the TSFZ geometry. Furthermore, the choice of feed rod composition does not appear to impact the crystallinity of the resulting h-BN. Figure S10 shows Raman data collected on a sample (\textbf{9}) grown under similar conditions as sample \textbf{2}, but with a B feed rod rather than a h-BN feed rod. The FWHM and $x_{0}$ of the modes are similar to what is observed for sample \textbf{2} (Fig. \ref{fig5}b) indicating similar quality by this metric. However, there are still tradeoffs with respect to the dynamics of the growth process itself. A B feed provides high zone stability due to superior wettability by the flux and allows faster \textit{v}$_{feed}$ at the same laser power due to the higher dissolution rate. TSFZ growth with a B feed rod also favors a proportionally larger polycrystalline h-BN seed rod size to maintain equalized B mass transfer, leading to a larger crystal. Recrystallization of the B rod (i.e. FZ growth) prior to usage in the h-BN growth may also improve the precursor purity as compared to the polycrystalline h-BN rods. However, even after recrystallization, some issues with thermal cracking remain, which can destabilize the growth. Furthermore, as mentioned previously, under certain conditions ($P$ = 5 – 7 bar, Fe or FeB$_{0.2}$ flux) usage of a B feed in the TSFZ geometry led to recrystallization of primarily polycrystalline h-BN and growth of dendritic, crystalline flakes. We suspect that this behavior arises due to a rapid supersaturation of the flux with B and N which leads to consitutional supercooling at the growth front and an unstable interface. This is possibly supported by the observation of the spontaneous nucleation of h-BN "islands" on the flux surface, as seen in Fig. S1c and Fig. S1d. 

In contrast, a h-BN feed rod is robust against thermal cracking, although its wettability is worse, leading to more zone stability issues, and a slower dissolution rate. These zone stability issues are somewhat ameliorated by using a smaller diameter (3 mm) polycrystalline h-BN feed rod, however this requires use of a matching smaller diameter polycrystalline h-BN seed rod to equalize the mass transfer of B which leads to a smaller grown crystal. 

For the LPG geometry, the B feed rod is essential. For sample \textbf{6} grown in the LPG geometry with a h-BN feed ($D_{L}$ = 6 mm) and a FeCrB$_{0.2}$ flux (Fig. S1b), the initial region is largely flux-free h-BN (see microCT data in Fig. S2b), however the remainder of the growth is predominately flux. The cross-over likely occurred once the initial presaturation of B in the flux was recrystallized as h-BN, since little of the h-BN feed rod dissolved. In contrast, using a $D_L$ = 6 mm B crystal as the feed rod, under otherwise identical conditions, resulted in extended h-BN recrystallization for sample \textbf{10} (see Fig. S1f and Fig. S2f). Regardless of growth geometry, the successful growth of h-BN crystals using a B feed rod indicates that, as in CBF-growth \cite{liu2018,li2020a}, sufficient N for the reaction can be supplied via the gas environment. An important implication of this result is that the isotope composition of crystals grown by this technique can be controlled. Isotope control, as has been reported previously \cite{yuan2019,giles2018,vuong2018}, has important consequences for properties optimization. 

We now summarize the currently optimized growth conditions. Here, optimization is parameterized by the amount of trapped flux as indicated by the microCT data and the crystal quality as measured by Raman and PL spectroscopies. The currently optimized configuration based on these parameters is growth with the TSFZ geometry using an Fe flux and polycrystalline h-BN rods with \textit{v}$_{growth}$ = 0.1 mm/h and $P$ = 5 – 7 bar ($P_{N2}  = $ 4.5 – 6.3 bar, with the given gas flow rates). This configuration yielded large volumes of crystalline h-BN with minimal flux inclusions and high-quality indicated by the spectroscopic characterization. 

The quality of our best TSFZ-grown sample (\textbf{1}) can be benchmarked against crystals grown by the CBF techniques by comparison of the PL and Raman spectra. In Table \ref{table1} we show a comparison of quality metrics from the spectroscopic data for both variants of CBF growth and sample \textbf{1}, as well as a dendritic flake from sample \textbf{8}. The average FWHM of the Raman peaks for sample \textbf{1} is essentially the same as the reported values for a sample grown by the APHT CBF technique from an Fe flux \cite{li2021}, which are both quite close to the best value reported for crystals from the HPHT CBF technique. In Fig. \ref{fig6}b we make a direct comparison of PL spectra for sample \textbf{1} and CBF samples grown from the APHT technique. The peaks corresponding to the phonon-assisted free exciton modes are comparable, with similar linewidths and relative intensities, indicating similar quality by this metric. Note that the shift in energy of these features for the isotope enriched CBF sample is due to the change in atomic masses, which affects the energies of the phonons involved. 

On the low-energy side, the C defect signal at 4.1 eV is significantly stronger ($\approx 5\times$) in the natural isotope abundance CBF sample than in sample \textbf{1}, qualitatively indicating that, with the same flux, the TSFZ method gives higher purity. In contrast, the CBF sample grown with isotopically enriched B and N shows no resolvable peak at 4.1 eV. We speculate that this improvement in purity may be due to a lower C content in the isotopically enriched B precursor although it may also be related to the difference in flux composition. Notably, there is a clear difference in the overall shape of the background in this energy range for the TSFZ and CBF samples: both CBF samples show a broad feature centered at roughly 4.3 eV, while sample \textbf{1} shows a broad feature centered at roughly 3.9 eV. These features occur together in measurements of commercially available h-BN powders, where they are superimposed on the sharp point defect related peaks at 3.9 eV and 4.1 eV \cite{bourrellier2016,museur2008a,museur2008b}. Prior work indicates that these broad features are related to complex DAP recombination mechanisms associated with native defects in h-BN (i.e. B or N vacancies/interstitials) \cite{bourrellier2016,museur2008a,museur2008b}. That the TSFZ-grown sample shows only the 3.9 eV feature while the CBF-grown samples show only the 4.3 eV feature may indicate that the two features result from distinct DAP recombination mechanisms and that the different growth techniques each favor only one type of the relevant defect. 

In the $E$ = 4.8 – 5.7 eV region we see that the TSFZ samples all show significantly higher PL intensity than the CBF ones, suggesting a higher density of the associated defects. This is reflected in the higher PL ratio shown in Table \ref{table1}. If this behavior is intrinsic it would suggest that an even slower growth rate ($v_{grow}$ < 0.1 mm/h) is needed to reduce the stacking fault density. However, we note that the dendritic flake samples (measured without any cutting or polishing) show a significantly lowered PL ratio, comparable to the CBF samples. Furthermore, the TSFZ samples measured in Fig. \ref{fig6} were sectioned out of larger boules and then mechanically polished prior to the PL measurement, unlike the CBF-grown samples, which provide suitable surfaces for PL measurements as-synthesized. Previous work has shown that the excitation spectrum of h-BN crystals changes markedly due to deformation induced by even minimal amounts of pressure \cite{watanabe2006}. Therefore, it is possible that the cutting and polishing process may result in an extrinsic increase in the stacking faults density causing a concomitant enhancement of the signal in this energy region. As a result, the intrinsic stacking fault density characteristic of the TSFZ technique is currently unclear and further PL measurements are in progress to address this point.

\begin{table}[]
\caption{\textrm{A comparison of crystal quality metrics from Raman and PL spectroscopies. For the APHT (Fe flux) and HPHT CBF techniques, the best (lowest) values found from the literature are shown.  For the TSFZ samples, results from sample \textbf{1} (Fig. \ref{fig5}a and Fig. \ref{fig6}a) and from the dendritic flakes found in sample \textbf{8} (see Fig. S1d, Fig. S2d and Fig. S5a) are shown. For the Raman data, we show the (average) FWHM of the intralayer E$_{2g}$ phonon mode. The variable "PL ratio" is defined as the intensity ratio PL(5.48 eV) / PL(5.77 eV), which correspond to features from stacking faults and the phonon-assisted free exciton luminescence, respectively. }}\label{table1}
\begin{tabular*}{\tblwidth}{@{}CC@{}CC@{}}
\bottomrule
  \textrm{\textbf{Technique}} & \textrm{\textbf{Raman FWHM [cm$^{-1}$]}} & \textrm{\textbf{PL ratio}} \\ 
\toprule
\textrm{CBF - APHT}  & \textrm{7.6 \cite{li2021}} & \textrm{2.4 (Fig. \ref{fig6}a), 1.1 \cite{li2021}} \\
\textrm{CBF - HPHT} & \textrm{7.3 \cite{schue2016}} & \textrm{0.9 \cite{watanabe2009}} \\
\textrm{TSFZ - sample \textbf{1}} & \textrm{7.7} & \textrm{20.9} \\
\textrm{TSFZ - sample \textbf{8}} & \textrm{7.6} & \textrm{2.7} \\
\bottomrule
\end{tabular*}
\end{table}

\section{Conclusions}
The intrinsic characteristics of the TSFZ technique make it well-suited to addressing several of the challenges associated with h-BN single crystal growth, namely increasing sample volume while maintaining high purity and crystallographic perfection. Here, we have shown the successful growth of h-BN crystals via this technique and thereby demonstrated its feasibility. As benchmarked by Raman and photoluminescence spectroscopies, our currently established optimal growth procedure produces bulk h-BN crystals on par with, or superior to, existing CBF-grown crystals in terms of purity and crystallinity. However, there remain many avenues for optimization of the TSFZ technique, as well as open questions. While the TSFZ samples reported here mostly show higher stacking fault densities than reported CBF samples, it remains to be determined whether this is intrinsic to the employed TSFZ growth parameters. Further explorations of the growth rate, increased growth pressure and flux composition are all desirable to understand how these variables interact to impact crystal quality and flux incorporation. Determining how changes in the rotation rate and thermal gradient affect the growth interface shape and grain nucleation and growth will also be helpful. Additionally, the effect of growth geometry should also be further explored via additional characterization of the LPG-grown samples. Lastly, it will be necessary to adapt current exfoliation methods to these crystals in order to incorporate them into heterostructures and determine their impact on device performance. Overall, these results have important implications for the successful implementation of the many technological and fundamental research applications of h-BN single crystals.

\section{Acknowledgements}

We thank Davor Tolj, Tiffany Soetojo and Rachel Michel for helpful discussions. This work made use of the bulk crystal growth facility of the National Science Foundation's Platform for the Accelerated Realization, Analysis, and Discovery of Interface Materials (PARADIM), which is supported by the National Science Foundation under Cooperative Agreement No. DMR-2039380. This work also utilized the JHU Raman Scattering Users Center. Work at UC Santa Barbara used facilities supported via the UC Santa Barbara National Science Foundation Quantum Foundry funded via the Q-AMASE-i program under award number DMR-1906325. This work was also supported by the BONASPES (ANR-19-CE30-0007) and HETERO-BNC (ANR-20-CE09-0014-02) projects. The MPMS was funded by the National Science Foundation, Division of Materials Research, Major Research Instrumentation Program, under award No. 1828490. GY, CN, and RH acknowledge support from National Science Foundation Grants No. DMR-2300640 and No. DMR-2104036. SDW and RG acknowledge support from the Air Force Office of Scientific Research under award No. FA9550-23-1-0042. JHE acknowledges support from the Office of Naval Research award No. N00014-22-1-2582.

\section{Data availability}
All data collected in PARADIM facilities to carry out this work will be available upon publication at https://doi.org/
doi.org/10.34863/211a-6d45. Other data is available upon reasonable request to the authors.
\printcredits

\bibliographystyle{cas-model2-names-noSort}


\end{document}